\documentclass[sigconf,balance=false]{acmart}

\setcopyright{none}
\renewcommand\footnotetextcopyrightpermission[1]{} 

\usepackage{tikz,graphics,color,fullpage,float,caption,subcaption}
\usepackage{multirow}
\graphicspath{{figures/}}
\usepackage{listings}
\usepackage{enumitem}
\usepackage{algorithm}
\usepackage{algpseudocode}
\usepackage{pifont}
\usepackage{geometry}
\copyrightyear{2026}
\acmYear{2026}
\setcopyright{cc}
\setcctype{by}
\acmConference[ICPP '26]{Proceedings of the 55th International Conference on Parallel Processing}{September 28-October 01, 2026}{Singapore, Singapore}
\acmBooktitle{Proceedings of the 55th International Conference on Parallel Processing (ICPP '26), September 28-October 01, 2026, Singapore, Singapore}
\acmDOI{10.1145/3832810.3832907}
\acmISBN{979-8-4007-2657-6/2026/09}

\begin{document}

\title[Fine-grained Computation-Communication Overlap for Mixture-of-Experts]{Fine-grained Computation-Communication Overlap via Tile-level Signaling and Scheduling for Mixture-of-Experts}


\author{Minyu Cui}
\orcid{1234-5678-9012}
\affiliation{%
  \institution{Linnaeus University }
  \streetaddress{Växjö, Sweden}
  \city{Växjö}
  \country{Sweden}
  \postcode{352 52}}
  \email{minyu.cui@lnu.se}

  \author{Anna Wingkvist}
\affiliation{%
  \institution{Linnaeus University}
  \city{Växjö}
  \country{Sweden}}
\email{anna.wingkvist@lnu.se}

\author{Morgan Ericsson}
\affiliation{%
  \institution{Linnaeus University}
  \city{Växjö}
  \country{Sweden}}
\email{morgan.ericsson@lnu.se}



\begin{abstract}
Mixture-of-Experts (MoE) architectures increase model capacity without proportionally increasing computation cost and have become a key building block for scaling
large language models (LLMs) to trillion-parameter regimes. 
Efficient deployment of these MoE models relies on distributed execution across multiple GPUs, where each MoE layer involves two all-to-all communications: dispatching tokens to expert ranks and returning the expert outputs to their source ranks. Conventional MoE implementations launch this return all-to-all after expert compute completes, exposing communication latency on the critical path and reducing GPU utilization.  
We present a fine-grained  approach that overlaps expert
compute with the second all-to-all via tile-level signaling and scheduling.  
Our producer-consumer co-design combines: (1) a persistent per-rank computation kernel (producer) that covers all local experts on the rank to eliminate repeated kernel launch overhead and prioritizes remote-critical tiles,  
and (2) a persistent communication kernel (consumer) on a small dedicated partition of streaming multiprocessors (SMs) that issues segment-granular transfers as tiles become ready. 
Our co-design avoids intrusive changes to the underlying computation operators or communication primitives, making it practical for improving distributed MoE execution efficiency on multi-GPU systems.
On a 4-A100 GPU platform, evaluated on three MoE models against four state-of-the-art MoE systems,  
our approach achieves up to 2.64x end-to-end speedup and 2.74x MoE-layer speedup.
Compared with a conventional non-overlap baseline, our approach consistently improves both operator- and MoE-layer-level performance across varying GEMM shapes, router modes, and a broad range of producer/consumer SM partitions, while preserving correctness.

\end{abstract}




\maketitle

\section{Introduction}

The computational demands of modern machine learning (ML) models are increasing rapidly, driven by growing model sizes, longer sequence lengths, and the rise of multimodal models.
Recent large language models have grown from billions to trillions of parameters in just a few years~\cite{fedus2022switch,llama4}.  
In dense transformer blocks, each token is processed by the same dense computation path, so per-token compute scales with model size. 
As models continue to grow, this coupling between model capacity and active per-token compute is becoming a primary scalability bottleneck.

The pressure to scale model capacity without proportionally increasing active computation has motivated sparse MoE architectures. 
The MoE layers route each token to only a subset of experts, thereby increasing model capacity while limiting the computation performed per token. 
However, in distributed MoE execution, the experts are typically partitioned across GPUs through expert parallelism, and token routing introduces substantial all-to-all communication overhead, accounting for nearly half of total execution time~\cite{Comet2025}.
This problem is further amplified by hardware trends: GPU compute throughput has scaled much faster than interconnect bandwidth across recent generations, causing accelerators to spend an increasing fraction of time idle during communication phase.

A natural way to mitigate this communication overhead is to overlap communication with the dependent computation. Several decomposition-based methods~\cite{ASPLOS2023_Wang,ASPLOS24_Chen,jangda2022breaking,Domino2024,Aync_TP,MegaScale_USENIX} split collectives and surrounding GEneral Matrix Multiplications (GEMMs) into chunks and pipeline them, typically in separate streams. MoE-specific systems~\cite{hwang2023tutel,he2022fastermoe,ScheMoE2024,PipeMoE_2023,Aync_TP,MPipeMoE_2023} apply the similar idea along the token dimension, pipelining experts compute with all-to-all communication. 
These coarse-grained overlap methods are straightforward to implement in existing frameworks, but they often do not fully exploit tensor-core efficiency at small chunk sizes and incur non-trivial host-side synchronization costs between chunks. 
In addition, collective APIs from GPU communication libraries such as NCCL usually require 
contiguous-buffer, so the decomposition is often limited to a single tensor dimension and the resulting chunks tend to be misaligned with the GEMM tile structure. 
As a result, tile-level overlap is hard to achieve through decomposition alone.
Another major bottleneck is the overhead of host-side kernel launches.
Kernel-fusion methods~\cite{Punniyamurthy_SC2024,FlashMoE,CCFuser2025,Comet2025} instead fuse computation and communication into a single kernel, thereby eliminating both host-side orchestration~\cite{Punniyamurthy_SC2024} and redundant kernel launches~\cite{FlashMoE}. 
These solutions universally improve overlap efficiency, but typically require intrusive engineering, including custom barriers, inter-rank atomic protocols, and per-target kernel specialization. 
Recent studies~\cite{eurosys26,T3_Pati_ASPLOS} use tile-level signaling to initiate communication from completed GEMM output regions, enabling finer-grained computation-communication overlap.
Our design adopts this signaling principle while keeping compute and communication kernels separate.

In this paper, we target distributed MoE inference and present a method that achieves fine-grained overlap by coordinating GEMM (producer) and communication (consumer) kernels through tile-level device-resident signals. Accordingly, our design and evaluation cover forward-pass execution only. Supporting the backward pass for training is left for future work.
The producer raises a signal when a tile completes, and the consumer issues a segment-granular transfer for one or more row bands (a row band is a contiguous strip of output rows spanning the full output width, defined in Section~\ref{sec:row_layout}, illustrated in Figure~\ref{segment_composition}) as soon as the corresponding tiles are ready, without host-side synchronization. 
Compared with decomposition-based methods, this design preserves tensor-core efficiency and incurs neither inter-chunk synchronization nor the kernel-launch overhead of chunked execution.
It also avoids the implementation complexity of fusion-based approaches.
Concretely, we launch both kernels as persistent kernels on disjoint SM partitions: a tunable share of SMs is dedicated to the communication kernel, while the remaining SMs run the GEMM. This partition gives each kernel interference-free compute resources, while persistent kernels and device-resident signals keep host and barrier synchronization off the critical path.
Realizing this design for the return path of an MoE layer, i.e.,  the expert output projection followed by the second all-to-all communication, is more challenging than for static communication patterns used in tensor parallelism. 
The difficulty arises because each expert output row must return to the rank that originally held its input token, and this row-to-rank mapping is determined at runtime. 
Our key advance is a remote-owner-aligned row layout that aligns every output tile with a single destination rank, so the consumer issues one contiguous remote write per segment, i.e., row band(s), without per-row routing logic inside the communication kernel~\cite{eurosys26}.

In summary, this paper makes the following contributions:
\begin{itemize}
\item \textbf{Fine-grained computation–communication overlap for MoE return path.} 
We propose a signaling-based, fine-grained overlap mechanism for the return path 
of distributed MoE layers. The GEMM and communication kernels run as persistent kernels on disjoint SM partitions, with the consumer assigned to a small, tunable subset of SMs.
The two kernels coordinate through tile-level device-resident signals. As soon as all tiles in a segment are signaled complete, the consumer issues that segment as one remote write, minimizing the gap between data production and transfer.
\item \textbf{Communication-aware remote-owner-aligned row layout.} 
We apply a remote-owner-aligned row layout that maps every output tile to a single destination rank. 
This layout allows the producer to prioritize tiles on the remote critical path and enables the consumer to issue each segment as one contiguous remote write. 
\item \textbf{Rank-wide GEMM kernel.}
A single persistent producer kernel per rank covers all local experts, eliminating repeated kernel-launch overhead and amortizing scheduling cost across the full set of GEMM tiles.
\item \textbf{Evaluation across MoE models and configurations.} 
We conduct a comprehensive evaluation across three MoE models, three routing distributions, varied GEMM shapes, and a broad range producer/consumer SM partitions, with  operator-, MoE-layer-, and end-to-end speedup analysis, SM-partition contention study, and correctness validation.

\end{itemize}

The paper is organized as follows. Section~\ref{background} reviews distributed MoE, GEMM tile signaling, and device-initiated communication. 
Section~\ref{our_approach_overlap} presents our producer-consumer co-design. 
Section~\ref{ER} evaluates our approach. 
Section~\ref{RW} discusses related work, and Section~\ref{Conclusion} concludes this paper.

\section{Background}
\label{background}

\subsection{Mixture-of-Experts Architectures}

In transformer-based models, an MoE layer replaces the standard feed-forward network (FFN) with multiple sub-layers (known as \textit{experts}) and a gating (router) network~\cite{lepikhin2021gshard}.  
For each token, the router selects the $topk$ experts via a probability distribution, and the layer output is the gate-weighted sum of their outputs. Because only a few experts run per token, the active compute is much smaller than that of a dense feed-forward network with the same parameter count. This sparsity is the property that lets MoE scale model capacity without proportionally scaling compute cost.

In distributed settings, experts are distributed across GPUs (or ranks).
A forward pass through the MoE layer requires:
\begin{enumerate}
\item \textbf{Routing:} Each rank evaluates the gating function on its local tokens to produce per-token expert assignments and gate weights.
\item \textbf{First all-to-all (dispatch):} Tokens are dispatched to the ranks hosting their assigned experts.
\item \textbf{Expert computation:} Each rank executes the expert computations for its local experts on the received tokens.
\item \textbf{Second all-to-all (combine):} Expert outputs are returned to each token's original rank after expert compute. 
\item \textbf{Weighted reduction (scale):} The owner rank scales its per-token output as a gate-weighted sum of the returned expert outputs.
\end{enumerate}
The two all-to-all operations are the primary communication bottleneck~\cite{Lancet2024}.
This paper focuses on overlapping expert compute with the second all-to-all.

\subsection{GEMM Kernels and Epilogue Signaling}

General matrix multiplication (GEMM), formulated as $C_{M\times N} = A_{M\times K} \times B_{K\times N}$, is the dominant computational operation in modern neural networks. 
Modern GPU GEMM kernels decompose the output matrix into rectangular output tiles.
A thread block, or cooperative thread array (CTA), typically computes one output tile at a time. 
It iterates along $K$ dimension, with its warps using tensor-core instructions to perform matrix multiply-accumulate operations, and accumulate partial sums in registers. After the reduction is complete,
the final results are written to global memory.
This tile structure is central to GPU GEMM performance. 
It reveals parallelism across the output matrix, improves data reuse through shared memory and registers, and provides a natural unit of scheduling. 
Crucially for our work, it also provides a natural granularity for producer-consumer coordination. Because CTAs are scheduled independently, individual output tiles become
ready before the full GEMM completes, allowing downstream work to consume ready tiles without waiting for the entire kernel to finish.

Libraries, such as CUTLASS, enable the tiled structure through templates, including tile shape, memory layout, pipeline depth, and epilogue operator.
A GEMM kernel consists of a main loop, performing the repeated tensor-core multiply-accumulate operations, followed by an epilogue.
The epilogue performs output conversion, optional element-wise operations, and the final
store of each completed tile. Since this store is where the tile becomes visible in memory, the epilogue is the natural place to publish a tile-level completion signal. 
With epilogue signaling, the producer sets a device-resident flag after storing each tile, and a consumer kernel can poll these flags to process ready tiles while other GEMM tiles are still being computed. The overhead is a small, fixed number of additional instructions per tile and does not perturb the main loop. 
Tile-level signaling therefore enables fine-grained producer-consumer coordination entirely on the GPU, withou host-side synchronization. This
mechanism forms the basis of our design.

\subsection{Device-Initiated Communication}

Distributed GPU workloads increasingly rely on device-initiated communication: transfers issued from inside a CUDA kernel rather than launched from the host. Libraries such as NVSHMEM expose one-sided put/get APIs over a symmetric heap, so that a kernel on one rank can write directly into symmetric GPU memory of another rank. 
This makes it possible to implement the all-to-all of an MoE layer as a single CUDA kernel that issues remote writes of tiles, rather than a chain of host-driven collective calls. We use this primitive to implement the second all-to-all of the MoE return path as a single consumer kernel driven by the per-tile signals.

\subsection{Concurrent Kernels and SM Partitioning}

CUDA streams allow multiple kernels to execute concurrently. However, these concurrent kernels share the available SM resources and contend for them by default. 
Launching each kernel as a persistent kernel (one with a fixed CTA count, where each CTA stays resident on its SM until the kernel exits) lets the two kernels run on disjoint SM subsets, removing this SM-level interference. We use this mechanism to run the communication consumer on a small SM partition alongside the producer GEMM, detailed in Section~\ref{pro_com_con_schedule}.

\section{System Design}\label{our_approach_overlap}

\subsection{Overview}
\label{sec:design_overview}

\begin{figure}[t]
\centering
{\includegraphics[width=0.45\textwidth]{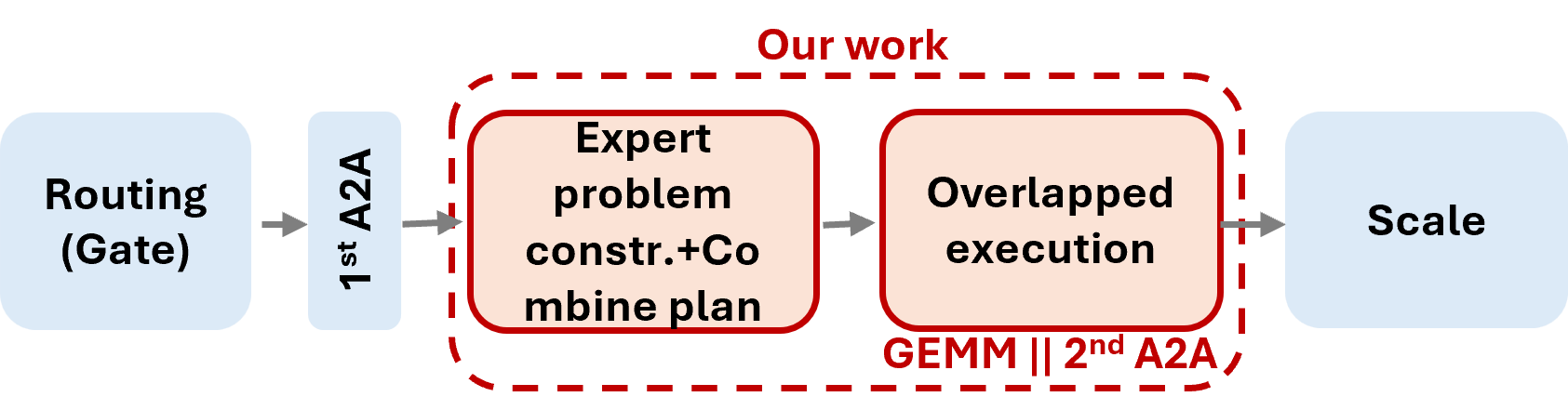}
\label{End_to_end_moe_layer_pipeline}}
\caption{Overview of our design for the MoE layer.}
\label{overview_our_design}
\end{figure}

\begin{figure}[t]
\centering
\begin{subfigure}[b]{0.9\linewidth}
{\includegraphics[width=1\textwidth]{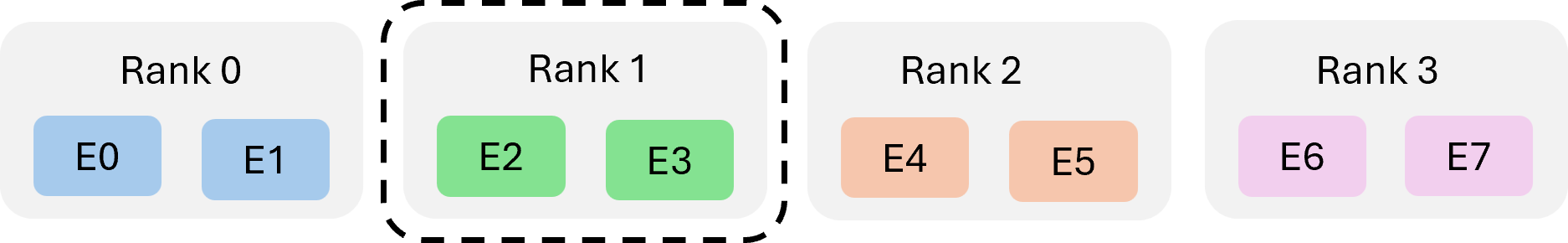}
\caption{4 ranks (R), 2 experts (E) per rank}
\label{4ranks_2experts}}
\end{subfigure}
\begin{subfigure}[b]{0.9\linewidth}
{\includegraphics[width=1\textwidth]{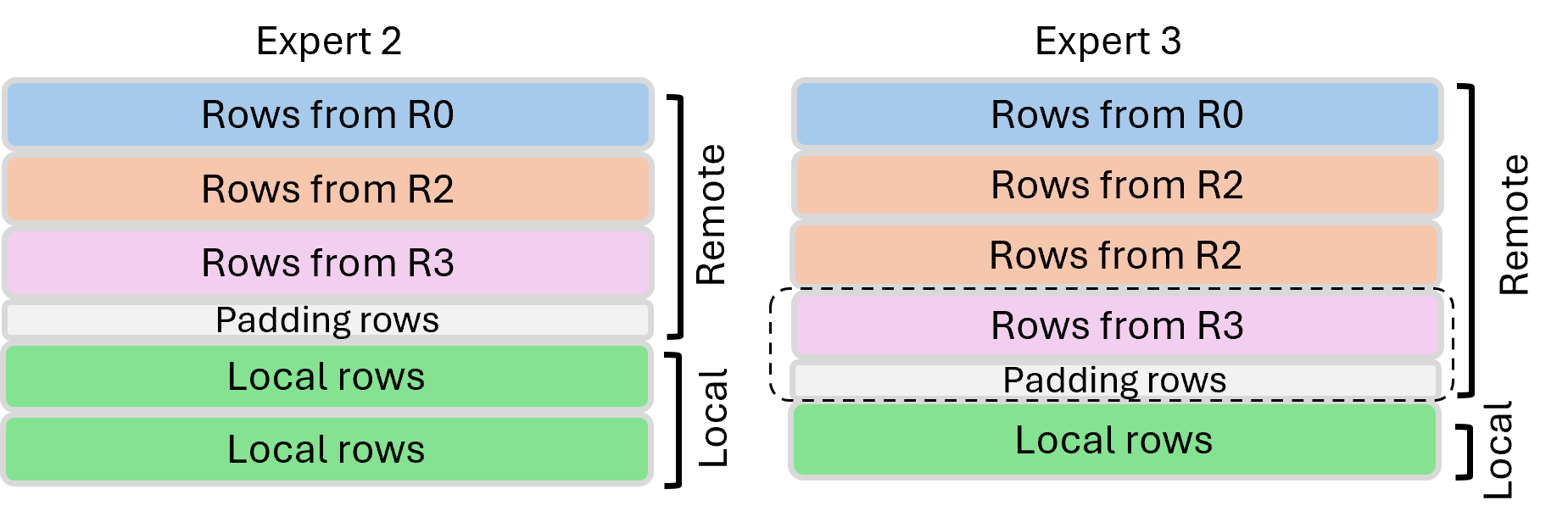}
\caption{Remote-owner-aligned row layout}
\label{GEMM_Output_Tiles_rank}}
\end{subfigure}
\begin{subfigure}[b]{0.9\linewidth}
{\includegraphics[width=1\textwidth]{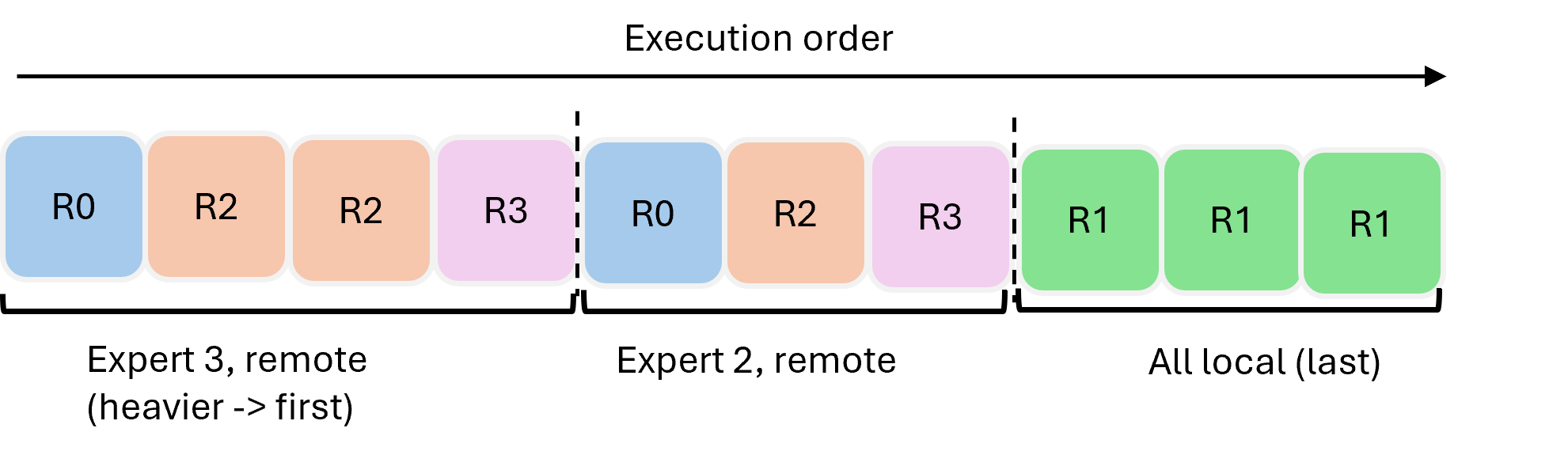}
\caption{Rank-wide tile schedule}
\label{Rank_wide_Tile_schedule}}
\end{subfigure}
\begin{subfigure}[b]{1\linewidth}
{\includegraphics[width=1\textwidth]{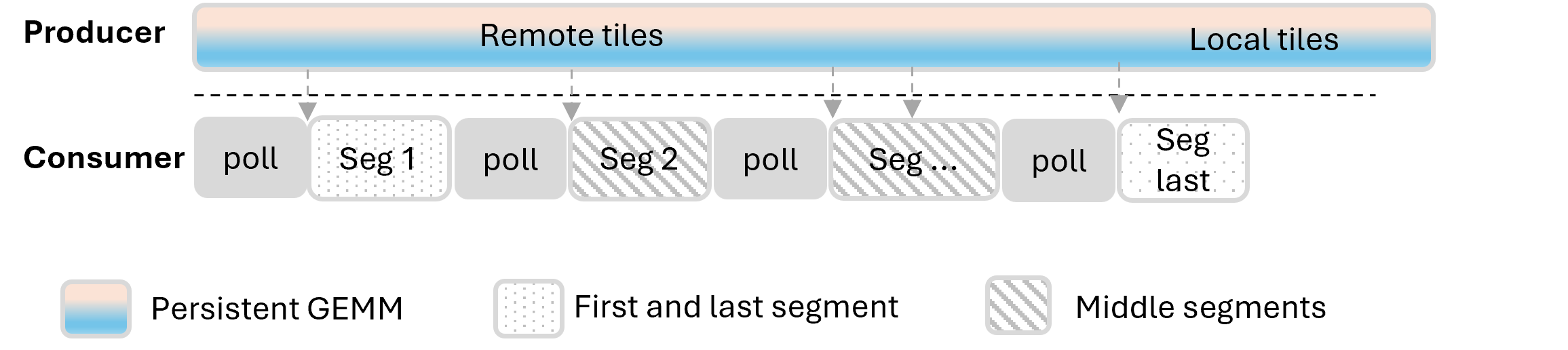}
\caption{Overlapped GEMM and second all-to-all}
\label{Overlapped_GEMM_2ndA2A}}
\end{subfigure}
\caption{Optimization strategies to hide communication.}
\label{Optimization_strategies}
\vspace{-6pt}
\end{figure}

Figures~\ref{overview_our_design}
and~\ref{Optimization_strategies} 
illustrate our design of the distributed MoE execution flow to enable fine-grained computation-communication overlap. 
We focus on the second all-to-all (return communication) after expert compute, which stays on the critical path in conventional MoE implementations. Throughout this section, each \textit{row} in the GEMM input corresponds to one routed token \textit{copy}; we use the two terms interchangeably. 
Our design is built around the following cooperating components:

\begin{enumerate}[label=\arabic*,]

\item \textbf{Expert problem construction and combine plan.} 
We impose a ``remote tokens first, local tokens last'' layout on the rank-wide GEMM input so that remote rows precede local rows and are grouped by destination (dest.) rank, with alignment padding inserted at owner-rank boundaries. We construct a tile schedule that produces remote-bound output earliest while minimizing expert transitions. We build a combine plan that pre-resolves all routing-derived transfer metadata into flat, per-tile device-resident arrays.
\item \textbf{Overlapped execution.} 
On the producer side, each rank launches a single long-lived persistent GEMM kernel that executes all local experts on the rank, and marks each completed output tile as ready through a lightweight signaling epilogue. On the communication side, a persistent consumer kernel runs concurrently on a disjoint subset of SMs, polls for ready data segments, and transfer them to their owner ranks via non-blocking NVSHMEM puts. 
The two kernels are launched on separate CUDA streams, with the consumer assigned higher priority, and coordinate entirely through device-resident flags with no host involvement on the critical path. 

\end{enumerate}

We detail each component in the following subsections.

\subsection{Expert Problem Construction and Combine Plan}

This phase, depicted by the left box of our work in Figure~\ref{overview_our_design}, prepares the input data, the execution order, and the transfer metadata for the rank-wide persistent GEMM kernel. 
Each expert retains its own $A$, $B$, and $C$ matrices, the rank-wide nature is achieved through a unified tile worklist rather than from matrix concatenation. 
The following three coupled mechanisms enable this design.

\subsubsection{Remote-owner-aligned Row Layout}
\label{sec:row_layout}

Fine-grained overlap is effective only if each completed output tile maps to a single destination rank. 
This is not automatically true, because after the first all-to-all, each local expert's input rows originate from multiple peer ranks. 
Without explicit reordering, a single output tile may contain rows assigned to different ranks, breaking the single-destination property required for fine-grained overlap. 
We impose a remote-owner-aligned row layout on the GEMM input. As shown in Figure~\ref{GEMM_Output_Tiles_rank}, for each local expert, we group rows by their original owner rank and order them such that:

\begin{enumerate}
\item 
Rows destined for \textit{remote} owner ranks appear first, grouped contiguously by owner rank;
\item 
Rows destined for the \textit{local} rank (i.e., tokens that originate on this same rank) appear last;
\item 
At each owner-rank boundary, if the group's row count is not a multiple of the tile dimension in $M$ (denoted $tb_M$), we pad the shortfall with zero-filled rows so that the next group starts at a tile-aligned offset. These padded rows participate in the GEMM computation for correctness but are excluded from the transfer. Figure~\ref{GEMM_Output_Tiles_rank} illustrates this with an example for the row group originally from rank 3, under a 4-rank, 2 experts-per-rank configuration. 
Padding only affects the trailing rows (those fewer than $tb_M$ rows) at each owner-rank boundary 
that would otherwise leave the next group misaligned. Full $tb_M$-height rows (defined as a \textit{row band}, illustrated in Figure~\ref{segment_composition}) within a group require no padding, and the local row groups require none. 
With this layout in place, each remote tile carries a statically known destination (dest.) rank, and the lookup from tile index to destination becomes trivial.
\end{enumerate}

\subsubsection{Remote-first Tile Schedule\label{tile_schedule}}
Given the aligned row layout, we construct a rank-wide tile schedule governed by two rules: (i) all remote tiles are produced before any local tiles, because only remote tiles require communication, and (ii) among experts, those with more remote rows are scheduled first, so that the largest transfers enter the communication stage earliest and expert transitions are minimized by completing each expert's remote tiles contiguously before moving to the next.
Figure~\ref{Rank_wide_Tile_schedule} shows the resulting scheduling for rank 1: expert 3's remote tiles run first (heavier remote load), followed by expert 2's remote tiles, and finally all local tiles. 
This scheduling maximizes the time window between the first transferable data segment and the end of the GEMM, giving the consumer the longest possible overlap to drain remote transfers.
Algorithm~\ref{alg:static-plan} summarizes the construction of the \textit{remote-owner-aligned} row layout and the \textit{tile schedule} on each rank.

\begin{algorithm}[t]
\caption{Remote-owner-aligned Row Layout and Remote-first Tile Schedule Construction}
\label{alg:static-plan}
\begin{algorithmic}[1]
\Require routing; local rank $r_\text{self}$; rank count $R$; Experts per rank $E$; tile-M dimension $tb_M$.
\Ensure per-expert row layouts $\{X_e, e\in E\}$; tile schedule  $S$.
\Statex \textit{Remote-owner-aligned Row Layout:}
\For{each local expert $e$}
  \For{each remote rank $r$ ($r \in \{0, \ldots, R-1\},\ r \neq r_\text{self}$)}
    \State $G_r \gets$ rows from rank $r$ routed to $e$.
    \If{$(|G_r| \bmod  tb_M) \neq 0$}
      \State pad $G_r$ with zero rows up to the next $tb_M$-multiple.
    \EndIf
  \EndFor
  \State append $G_0, G_1, \ldots, G_{R-1}$ (remote first, $G_{r_\text{self}}$ last) to $X_e$.
\EndFor
\Statex \textit{Remote-first Tile Schedule:}
\State sort local experts by remote row-band count in descending order. \Comment{rule (ii)}
\State $S \gets [\,]$
\For{each $e$ in sorted order} 
  \For{each row band $b$ in $X_e$ with $\text{dest.} \neq r_\text{self}$} 
    \State append tiles of $b$ to $S$.
    \Comment{rule (i): all remote tiles first}
  \EndFor
\EndFor
\For{each local expert $e$}
  \For{each row band $b$ in $X_e$ with $\text{dest.} = r_\text{self}$}
    \State append tiles of $b$ to $S$.
  \EndFor
\EndFor
\State \Return $\{X_e, e\in E\}; S$.
\end{algorithmic}
\end{algorithm}

\subsubsection{Combine Plan\label{sec:combine_plan}}

To return each tile to its owner rank, the consumer must know three things: the destination rank, the write offset in the remote receive buffer, and the number of valid (non-padding) rows. 
Resolving these at transfer time would require the consumer to inspect each tile's rows and look up their original owner ranks from the routing metadata produced in the gate stage, adding per-row overhead to every transfer. 
We avoid this cost with a combine plan that walks the row layout described earlier and constructs flat, per-tile device-resident arrays: destination rank, remote buffer offset, and valid row count. These arrays are indexed by tile identifier, reducing each tile's transfer resolution to three array reads.
Under the remote-owner-aligned layout, every tile's valid rows share a single owner rank, so this plan classifies each tile as either remote-publishable (single remote owner rank) or local-only. 
For each remote-publishable row band, the plan assigns a global write offset in the destination rank's receive buffer, coordinating placement across all local experts so that data from different experts targeting the same peer rank lands at non-overlapping offsets. Local-only tiles require no remote transfer.

\subsection{Overlapped Execution\label{overlapped_execution}}

This phase, depicted by the right box of our work in Figure~\ref{overview_our_design}, runs the producer and the consumer concurrently, as shown in Figure~\ref{Overlapped_GEMM_2ndA2A}. 

\subsubsection{Rank-Wide Persistent GEMM Kernel with Tile-Level Signaling}
\label{sec:rank_wide_gemm}

Rather than launching one GEMM kernel per expert, the producer is a single persistent GEMM kernel per rank that processes all tiles following the rank-wide tile schedule (Section \ref{tile_schedule}). 
Each scheduled item pairs an expert index with a tile coordinate, and a per-expert device view supplies the expert's $A$, $B$, and $C$ pointers along with its problem dimensions. This lets one kernel cover multiple experts without concatenating their matrices, distinguishing our approach from classical grouped GEMM.
The kernel is launched with multiple persistent CTAs, one per SM. Each CTA processes scheduled tiles in a strided pattern: CTA $i$ handles tiles $i, i{+}g, i{+}2g, \ldots$, where $g$ is the grid size.
For each tile, the CTA fetches the corresponding device view, copies the kernel's base parameter block, and overwrites only the expert-varying fields (matrix pointers and problem size) before invoking the GEMM tile routine. 
Because the substitution is a register-level copy rather than a re-launch, transitioning between experts is negligible. Shared-memory staging buffers are likewise reused across tiles regardless of expert identity.

Signaling is implemented inside the CUTLASS epilogue of the GEMM kernel, which runs after each tile's main-loop multiply-accumulate finishes and does not interfere with the GEMM's main loop.
After a CTA finishes writing an output tile to global memory, it performs a thread-fence to ensure that the data is globally visible, then publishes a readiness flag for the completed tile. 
This signaling mechanism forms the interface between the concurrent executions of the GEMM producer and the communication consumer.

\subsubsection{Communication Granularity}\label{sec:comm_granularity}

\begin{figure}[t]
\centering
\begin{subfigure}[b]{0.45\linewidth}
{\includegraphics[width=1\textwidth]{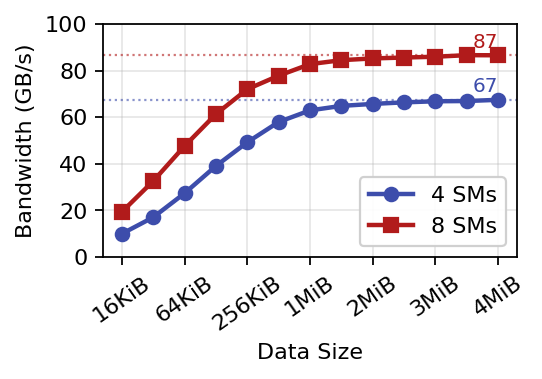}
\caption{Bandwidth curve}
\label{nvshmem_put_block_bw_curve_runtime}}
\end{subfigure}
\begin{subfigure}[b]{0.45\linewidth}
{\includegraphics[width=1\textwidth]{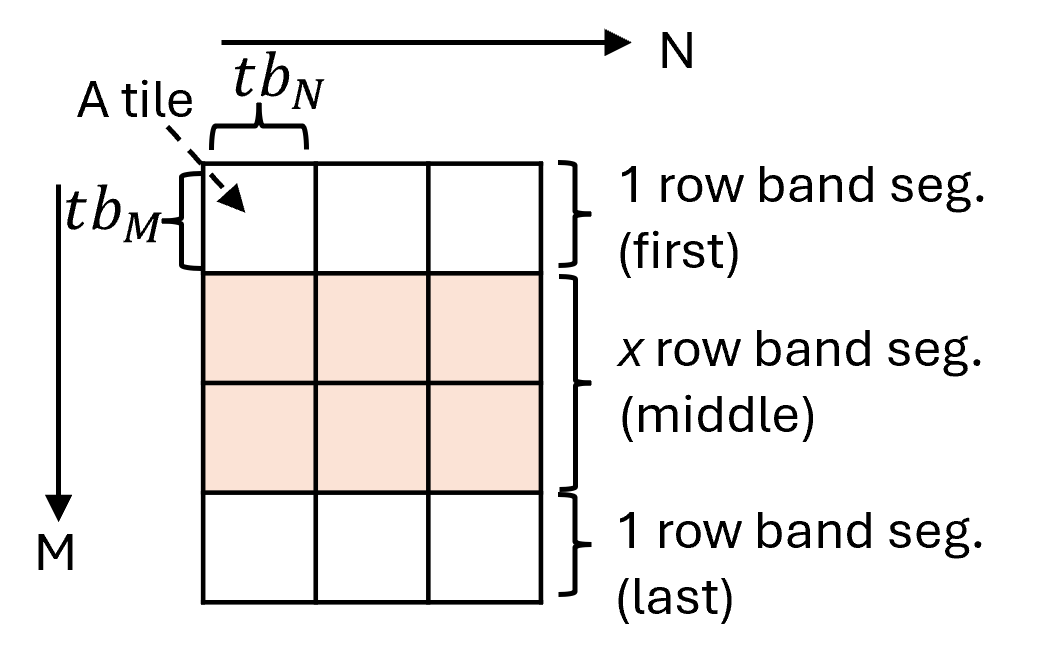}
\vspace{-6pt}
\caption{Segment composition}
\label{segment_composition}}
\end{subfigure}
\caption{Communication Granularity.  (a) Measured bandwidth varying with data size at two consumer SM budgets. Measurements are collected on the platform described in Section~\ref{platform}. (b) Segment composition: a transfer segment spans one or more $tb_M$-height row bands across the full output width $N$; first and last segments have one band, middle segments have $x$ tunable bands.}
\label{comm_granularity}
\end{figure}

The signaling mechanism described above marks each completed tile as ready, but transferring data per tile would yield significant communication fragmentation: the latency of tile-by-tile transfers becomes a non-trivial bottleneck~\cite{eurosys26}. 
The appropriate transfer granularity is determined by two factors: bandwidth utilization and the structure of the GEMM output under our row layout.

Figure~\ref{nvshmem_put_block_bw_curve_runtime} shows this effect using bandwidth curves varying with data size for two example consumer budgets (8 and 4 SMs, one CTA per SM).
The persistent consumer kernel occupies a fixed subset of SMs disjoint from the producer. A smaller consumer budget frees more SMs for the GEMM, while a larger budget achieves higher communication bandwidth up to a saturation point beyond which additional SMs usually offer no further bandwidth gain. 
At small data sizes, the curves sit in the low utilization regime, then bandwidth rises with data size, approaching the 87 GB/s peak at 8 SMs around 1 MiB and 67 GB/s at 4 SMs around 3 MiB. 
A single-tile transfer (usually 32 KiB or 64 KiB depending on the tile configuration used) lies far from saturation. This motivates transferring data at a granularity larger than a single tile.
Under the aligned row layout of Section~\ref{sec:row_layout}, all rows within one $tb_M$-height row band return to the same destination rank, making the entire row band the natural unit of transfer. 
An individual tile, in contrast, covers only $tb_N$ of the $N$ columns. Although such a tile is owner-uniform, its payload occupies only a partial-width slice of the row band and therefore appears as a strided, non-contiguous memory region, requiring multiple copies to transfer. 
A row band, however, spans the entire width $N$ and is both owner-uniform and contiguous in memory, so it can be moved with a single remote write. 

For the above reasons, we transfer data at segment granularity rather than individual tiles. 
The segment size involves a trade-off: small segments can start transfer sooner but achieve lower bandwidth, while large segments achieve higher bandwidth but delay the start of each transfer. As shown in Figure~\ref{segment_composition}, we resolve it with an tunable segment partitioning strategy: the first segment is kept at a single row band so that the consumer can issue its first transfer as soon as the row band is ready. 
The last segment is also a single row band to reduce the risk of a long communication tail: a larger final segment would have to wait for more tiles to complete before it could be issued, and would then take longer to transfer, both of which push communication past the end of the GEMM. 
The middle transfer segments coalesce $x$ consecutive row bands, where $x$ is a tunable parameter so that each transfer's payload falls on the saturated portion of the bandwidth curve in Figure~\ref{nvshmem_put_block_bw_curve_runtime}. 
The optimal $x$ is empirical and workload-dependent: our evaluation in Section~\ref{ER} shows that neither $x = 1$ ($\sim$1–2 MiB) nor $x = 2$ ($\sim$2–4 MiB) is uniformly best,
because merging more row bands into one interior segment pays off only when the resulting bandwidth gain exceeds the extra wait it imposes on the consumer.

\subsubsection{Producer-Consumer Co-Scheduling}\label{pro_com_con_schedule}

Sections~\ref{sec:rank_wide_gemm} and~\ref{sec:comm_granularity} define what the producer publishes and at what granularity the consumer transfers data. What remains is how the two kernels share the GPU so that GEMM progress and data transfer genuinely run in parallel. 
We implement producer and consumer as two independent persistent kernels and rely on three techniques: stream-level priority, SM-level partitioning, and device-memory coordination.

The two kernels run on separate non-blocking CUDA streams. We assign the communication kernel higher priority than the GEMM kernel since delaying a producer tile affects performance only at the GEMM tail, whereas delaying a consumer transfer pushes directly onto the communication critical path and compounds across subsequent segments.
The priority asymmetry also shapes how the SMs are shared between the two kernels.  Because the consumer kernel runs at higher priority, the SMs it occupies are effectively reserved for communication, leaving the remaining SMs available to the GEMM computation.
This removes SM-level interference from the hot path: the producer uses a stable SM budget throughout its execution, and the consumer has the SMs it needs to sustain the bandwidth plateau of Figure~\ref{nvshmem_put_block_bw_curve_runtime}. 
The consumer's SM budget itself is a tunable, coupled with the interior segment size. 
If the consumer budget is too small, the transfer pipeline falls behind producer progress and communication spills past the GEMM tail. If it is too large, the producer loses too much compute capacity, which lengthens GEMM and can offset the benefit of overlap. Section~\ref{ER} quantifies this tradeoff.
Once both kernels are resident, they coordinate entirely through device-resident state, without host intervention on the critical path.

Overall, our approach transforms the conventional serialized compute-then-communicate execution pattern into a continuously overlapped execution. Computation and communication progress concurrently, with the compute path driving communication through lightweight device-resident ready-state updates.

\subsection{Implementation}
\label{sec:implementation}

We implement our design as a CUDA runtime exposed to PyTorch 2.6.0, built on CUDA 12.1, CUTLASS 3.9, and NVSHMEM 3.6.5. The persistent GEMM kernel is built on top of the CUTLASS templated GEMMs~\cite{cutlass}. It preserves CUTLASS’s optimized main-loop structure by selecting the optimal configuration for each target GEMM shape from a lookup table derived from CUTLASS profiler.
Following EVT~\cite{EVT2024}, the tile-level signaling is integrated into the GEMM epilogue, so that signaling is fused into the producer without perturbing the main-loop multiply-accumulate. 
The consumer is implemented as a separate persistent kernel whose communication is performed via NVSHMEM.

\section{Experimental Results}\label{ER}

\subsection{Experimental Setup} \label{Exp_setup}

\subsubsection{Multi-GPU Platform and Software.}\label{platform}
We run all experiments on a single node with four NVIDIA A100 GPUs connected over
intra-node NVLink. 
Each NVLink lane delivers 25 GB/s per direction, and a GPU-to-GPU pair is connected by 4 NVLink lanes, giving a one-way bandwidth ceiling of roughly 100 GB/s per peer. Each GPU has 108 SMs and 40 GB HBM.
Each test uses one rank per GPU. 
The software environments include CUDA 12.1, NCCL 2.29.3, and PyTorch 2.6.0. NCCL carries the first all-to-all (dispatch) and the framework baseline's communication path, while the NVSHMEM and CUTLASS roles are as described in Section~\ref{sec:implementation}.

\subsubsection{Baselines}

We compare our approach against four state-of-the-art (SoA) MoE systems: (a) FasterMoE~\cite{he2022fastermoe} customizes the all-to-all primitive and pipelines it with expert communication, (b) Megatron-CUTLASS~\cite{shoeybi2019megatron, nvidia_grouped_gemm} implements expert computation with CUTLASS GroupGEMM kernel, (c) Megatron-TE~\cite{shoeybi2019megatron, nvidia_te} uses CUTLASS NVIDIA's Transformer Engine,  and (d) Tutel~\cite{hwang2023tutel} provides an adaptive MoE runtime with a tunable pipelining degree and a hierarchical 2D all-to-all.

\subsubsection{Workloads and Evaluation Scenarios.}\label{workload_evaluation}

Each test performs several warmup iterations (discarded in our evaluation) followed by measured iterations. 
We first evaluate our approach against the four SoA approaches using three MoE models. Table~\ref{tab:model-configs} summarizes the evaluated models derived from widely used transformer architectures: M-GPT, M-BERT, and M-Trans-xl. 
Each model consists of 12 transformer blocks, where selected FFN layers are replaced with MoE layers following the replacement setting in~\cite{CCFuser2025}. Specifically, M-GPT replaces the FFN with an MoE layer in the 11th block, representing a sparse-replacement scenario. M-BERT uses MoE in the 2nd, 5th, 8th, and 11th blocks, following an evenly-spaced replacement pattern commonly adopted in MoE literature~\cite{fedus2022switch,lepikhin2021gshard}. M-Trans-xl~\cite{dai2019transformerxl} uses MoE in all 12 blocks, representing the extreme of dense-replacement. 
All models use $topk=2$ routing with 64 total experts distributed across 4 GPUs (16 experts per GPU). 
We also evaluate $E$ $\in$ \{4, 8, 16, 32, 64\} to isolate how the MoE layer execution time scales with the number of experts per rank.

We further study the effectiveness and scalability of our design by implementing our approach on top of a conventional PyTorch MoE implementation, and comparing it against a baseline that executes the expert compute and return communication sequentially with standard primitives by calling cuBLAS and NCCL APIs.
The two systems are otherwise identical, so the measured speedup reflects the overlap mechanism alone. As listed in Table~\ref{tbl:design_quality}, we vary four parameters: the token count per rank, the GEMM shape (output hidden dimension), the router mode, and the SM partition between the producer kernel and the consumer kernel. We also tune the row bands (denoted as $mgb$) for the middle transfer segments.

\begin{table}[t]
\centering
\small
\caption{MoE model configurations. 
The MoE Blk. column lists the transformer blocks where the FFN is replaced by an MoE layer.
All configurations use
$topk=2$, $E_{total}=64$ experts ($E=16$), and 1 rank per GPU. 
}
\label{tab:model-configs}
\resizebox{\columnwidth}{!}{%
\begin{tabular}{lccccccc}
\toprule
Model & Batch & Seq & Tokens(B$\times$S) & Dim. &E\_{total} & MoE Blk. \\
\midrule
M-GPT        & 8  & 1024 & 8192  & 768 &64 & \{11\}  \\
M-BERT & 32 & 512  & 16384  & 1024 &64 & \{2,5,8,11\}  \\
M-Trans-xl   & 16 & 512  & 8192  & 512  &64 & \{all\} \\
\bottomrule
\end{tabular}%
}
\end{table}

\begin{table}[t]
\centering
\small
\caption{MoE workload shape sweep with $topk=2$.}
\label{tbl:design_quality}
\begin{tabular}{ll}
\toprule
\multicolumn{2}{c}{\textbf{Problem-shape sweep}} \\
\midrule
tokens per rank ($tpr$)   & 8192, 16384 \\
$M$ (= 2$\times$tpr)& 16384, 32768 \\
$K$ (hidden)        & 2048 \\
$N$ (out hidden)    & 4096, \textbf{8192} \\
router mode        & balanced, moderate\_skew, stress\_skew \\
experts per rank ($E$)  & 4, 8, 16, 32, 64 \\
\midrule
\multicolumn{2}{c}{\textbf{Implementations}} \\
\midrule
\textit{baseline (base)} & default SMs for GEMM and all-to-all kernels \\
$cCTA$ in \textit{ours} & 2, 4, 6, 8, 10, 12, 14 (default), 16, 18, 20, 22, 24\\
$mgb$ in \textit{ours} &  1, 2 \\
\bottomrule
\end{tabular}
\end{table}

\begin{figure}[htbp]
    \centering
    \includegraphics[width=\columnwidth]{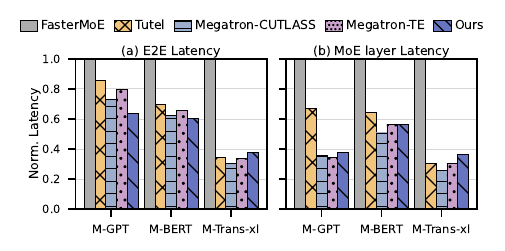}
    \caption{End-to-End and MoE layer latency ($\downarrow$ is better) in MoE models.}
    \label{ours_baseline}
\end{figure}

\begin{figure}[t]
\vspace{-5pt}
\centering
\begin{subfigure}[b]{0.23\textwidth}
{\includegraphics[width=1\textwidth]{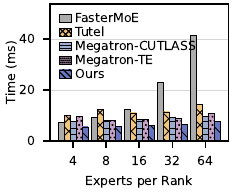}
\caption{Latency with varying E}
\label{moe_layer_runtime_vs_epr}}
\end{subfigure}
\begin{subfigure}[b]{0.23\textwidth}
{\includegraphics[width=1\textwidth]{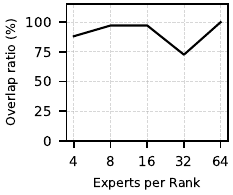}
\caption{Overlap ratio with varying E}
\label{overlap_ratio_vs_epr}}
\end{subfigure}
\caption{Single MoE-layer microbenchmark.}
\label{MoE_michrobechmark}
\end{figure}

\begin{figure}[t]
\centering
\begin{subfigure}[b]{1\linewidth}
{\includegraphics[width=1\textwidth]{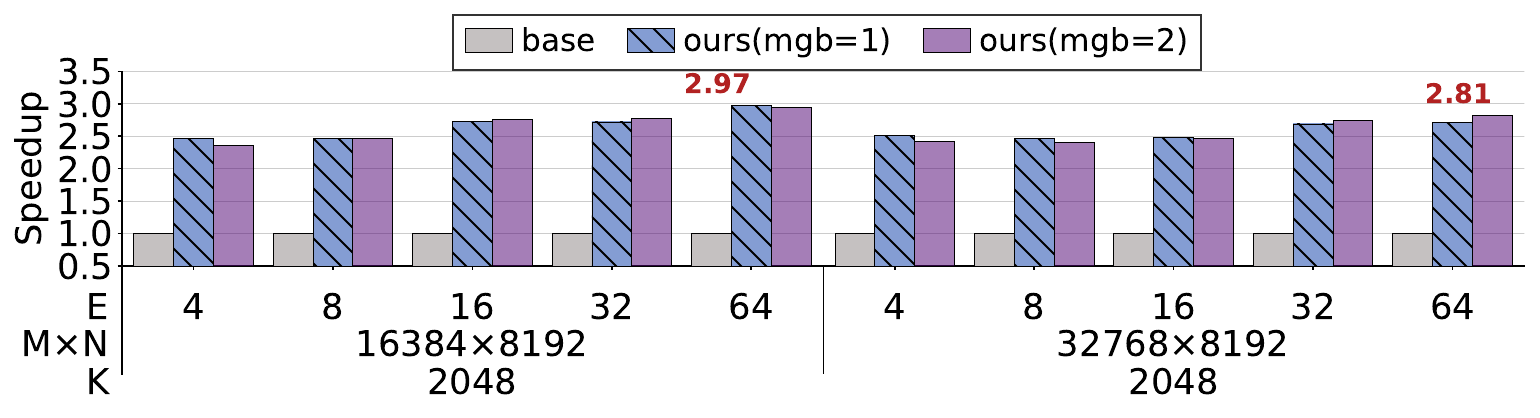}
\caption{balanced}
\label{speedup_vs_fw_avg_stage5_operator_max_ms_balanced_c8}}
\end{subfigure}
\begin{subfigure}[b]{1\linewidth}
{\includegraphics[width=1\textwidth]{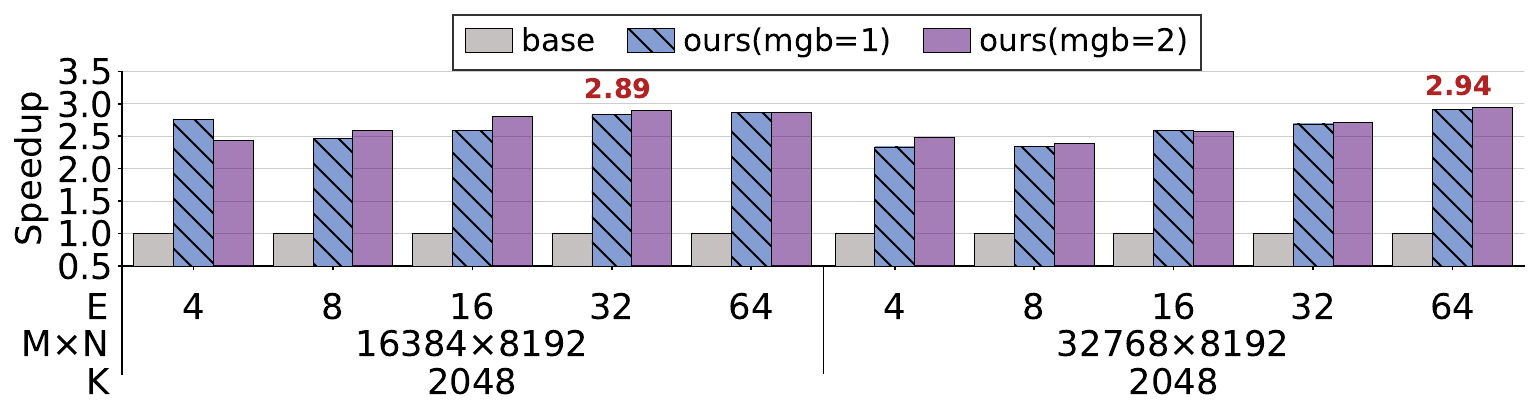}
\caption{moderate\_skew}
\label{speedup_vs_fw_avg_stage5_operator_max_ms_moderate_skew_c8}}
\end{subfigure}
\begin{subfigure}[b]{1\linewidth}
{\includegraphics[width=1\textwidth]{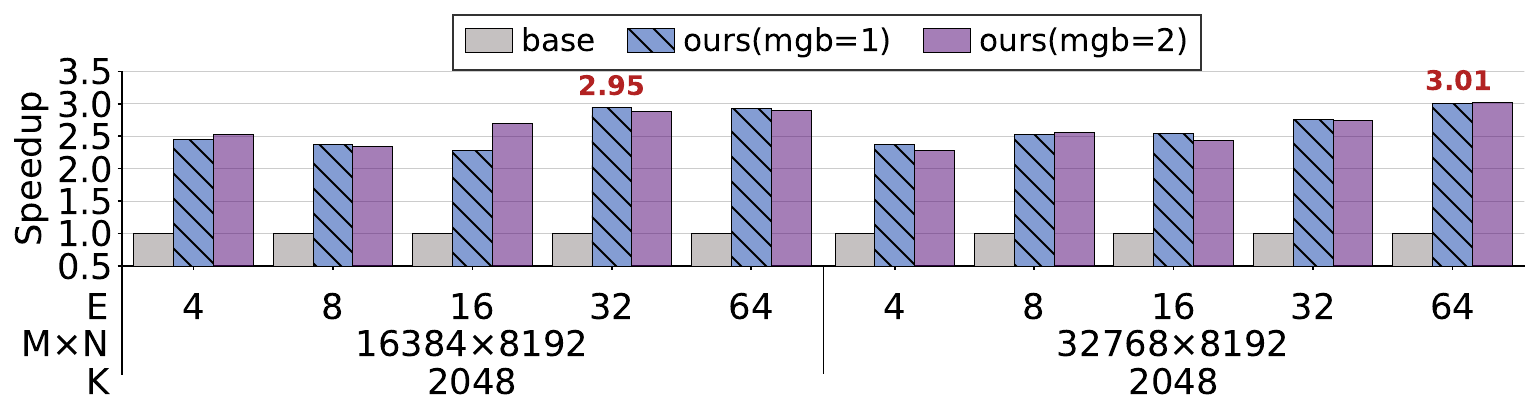}
\caption{stress\_skew}
\label{speedup_vs_fw_avg_stage5_operator_max_ms_stress_skew_c8}}
\end{subfigure}
\caption{Speedup at operator-level.}
\label{Operator_level_base_ours}
\end{figure}

\subsection{Performance Evaluation in MoE Models\label{sec:perf_moe}}

We report the proposed approach against four SoA MoE systems in two settings. Section~\ref{sec:e2e_moe_models} measures end-to-end forward latency and the MoE-layer latency on three production-style MoE models. Section~\ref{sec:moe_layer_E} uses a single MoE-layer microbenchmark to study how performance scales as the number of experts per rank varies.

\subsubsection{End-to-End and MoE Layer Evaluation in MoE Models}\label{sec:e2e_moe_models}

Figure~\ref{ours_baseline} depicts the end-to-end (E2E) and MoE-layer latency of FasterMoE, Tutel, Megatron-CUTLASS, Megatron-TE, and ours on the three MoE models, normalized to fasterMoE. 
For end-to-end latency in the left sub-figure of Figure~\ref{ours_baseline}, 
our approach consistently outperforms all four baselines for both M-GPT and M-BERT, with 1.57x and 1.66x speedups over FasterMoE, 1.35x and 1.15x over Tutel, 1.15x and 1.04x over Megatron-CUTLASS, and 1.25x and 1.09x over Megatron-TE. 
For M-Trans-xl, our approach delivers a 2.64x speedup over FasterMoE, 
but slightly trails Tutel, Megatron-CUTLASS, and Megatron-TE. 
This is likely because M-Trans-xl has the smallest hidden dimension ($Dim. = 512$) of the three models. Per-token all-to-all volume scales linearly with $Dim.$, so a smaller $Dim.$ leaves less communication to hide. 
At the same time, per-tile useful work is also reduced, so the fixed overhead of fine-grained overlap becomes a larger fraction of layer time.   
For the MoE-layer latency in the right sub-figure of Figure~\ref{ours_baseline}, we observe similar trends: 
our approach delivers decisive gains over the overlap-attempting baselines, with 2.65x and 1.78x speedups over FasterMoE and 1.77x and 1.15x over Tutel, on M-GPT and M-BERT. 
For M-Trans-xl, our approach outperforms FasterMoE by 2.74x, while remaining comparable, but does not surpass the performance of Tutel, Megatron-CUTLASS, and Megatron-TE. 

These results show that our proposed overlap mechanism remains effective after being integrated into full MoE model execution.
The two subfigures in Figure~\ref{ours_baseline} show that the end-to-end speedup latency is slightly lower than the corresponding MoE-layer speedup, 
especially for M-GPT and M-BERT.  
when comparing our approach with the other approaches.
This is expected: our method reduces only the time spent in the MoE layers, while the non-MoE components incur the same cost. 
By Amdahl's law, the end-to-end gain is therefore bounded by the MoE share of each model. 
Even under this bound, a substantial speedup is achieved, confirming that the MoE-layer improvement translates into practical end-to-end performance.

\subsubsection{MoE Layer Performance Across Expert Counts}\label{sec:moe_layer_E}

To characterize how each system scales with the expert count $E$, we vary $E$ from 4 to 64 on a single MoE-layer microbenchmark (1 MoE block, $B=16$, $S=512$, $topk=2$, $Dim=1024$ from~\cite{CCFuser2025}).
In Figure~\ref{moe_layer_runtime_vs_epr}, our approach achieves the lowest layer latency across the full range and outperforms all four baselines, with 1.30x to 5.33x speedups over FasterMoE, 1.77x to 2.16x over Tutel, 1.23x to 1.45x over Megatron-CUTLASS, and 1.35x to 1.76x over Megatron-TE. 
This scaling trend shows that the benefit of our overlap
increases with expert parallelism. With more experts per rank, both expert compute and return communication become heavier. 
In conventional designs, the added communication appears as a longer post-GEMM tail. Our producer-consumer co-design instead overlaps these transfers with ongoing GEMM execution, thereby reducing the exposed communication cost.
Figure~\ref{overlap_ratio_vs_epr} further supports this interpretation through the overlap ratio (the fraction
of the second all-to-all hidden behind the expert compute) of our approach. The overlap ratio stays consistently high throughout the full $E$ sweep, ranging from 71.9\% to 99.9\%, with several configurations approaching full hiding. This confirms that tile-level signaling exposes communication work early enough for the consumer to keep pace with the producer. 

In contrast to the production models in Section~\ref{sec:e2e_moe_models}, our approach now also achieves lower MoE-layer latency than Megatron-CUTLASS and Megatron-TE. This is potentially because the larger hidden dimension ($Dim. = 1024$) provides enough per-tile useful work to amortize the fixed overhead of fine-grained overlap.

\subsection{Scalability Analysis}

We have demonstrated both the MoE layer and the end-to-end effectiveness of our approach in Section~\ref{sec:perf_moe}. This section further analyzes its scalability under the configurations listed in Table~\ref{tbl:design_quality}.
Due to page limitation, we report only the configurations highlighted in bold in Table~\ref{tbl:design_quality} ($N=8192$). Similar trends hold for the remaining configurations unless explicitly noted. 
We adopt $cCTA=14$ by default, the other SM partitions are used only in our targeted resource contention experiments in Section~\ref{resource_contention}.

\subsubsection{Operator-Level Performance}
Figure~\ref{Operator_level_base_ours} presents the operator-level speedup (only considering expert GEMM plus second all-to-all) of our approach over the baseline.
Our method achieves higher speedup on all tests, reaching peak values of 2.97x under \texttt{balanced} routing, 2.94x under \texttt{moderate\_skew}, and 3.01x under \texttt{stress\_skew}. 
The speedup generally grows with $E$: larger $E$ amplifies the return communication, leaving more slack for the overlap mechanism to absorb. 

\begin{figure}[t]
\centering
\begin{subfigure}[b]{1\linewidth}
{\includegraphics[width=1\textwidth]{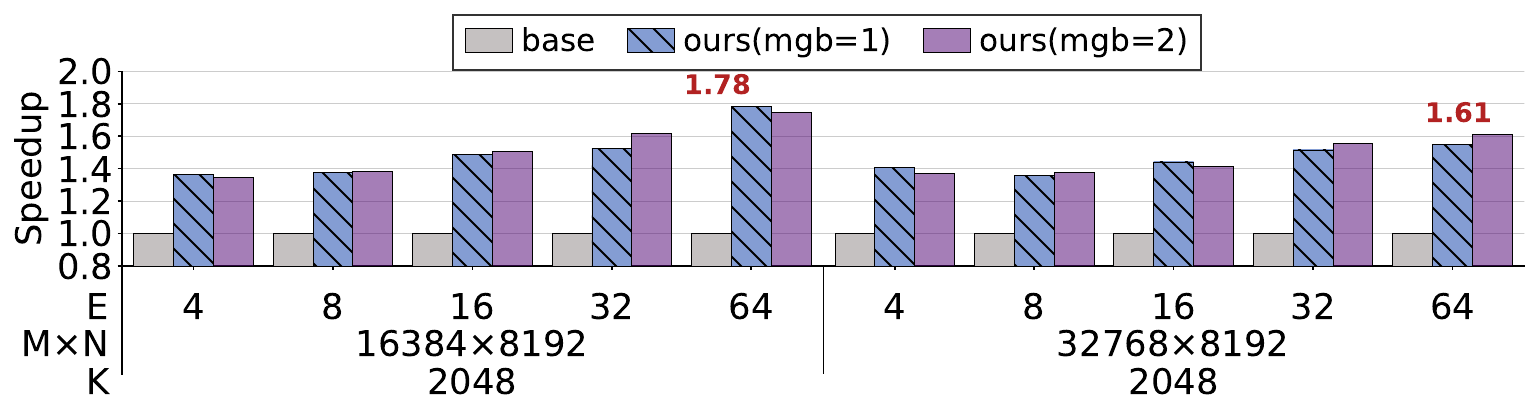}
\caption{balanced}
\label{speedup_vs_fw_avg_total_max_ms_balanced_c8}}
\end{subfigure}
\begin{subfigure}[b]{1\linewidth}
{\includegraphics[width=1\textwidth]{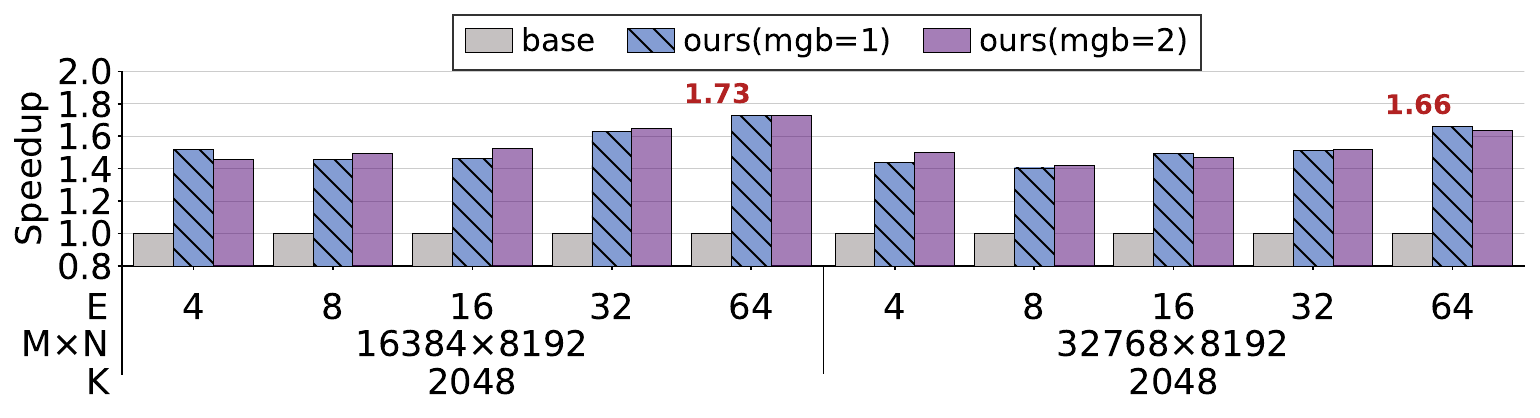}
\caption{moderate\_skew}
\label{speedup_vs_fw_avg_total_max_ms_moderate_skew_c8}}
\end{subfigure}
\begin{subfigure}[b]{1\linewidth}
{\includegraphics[width=1\textwidth]{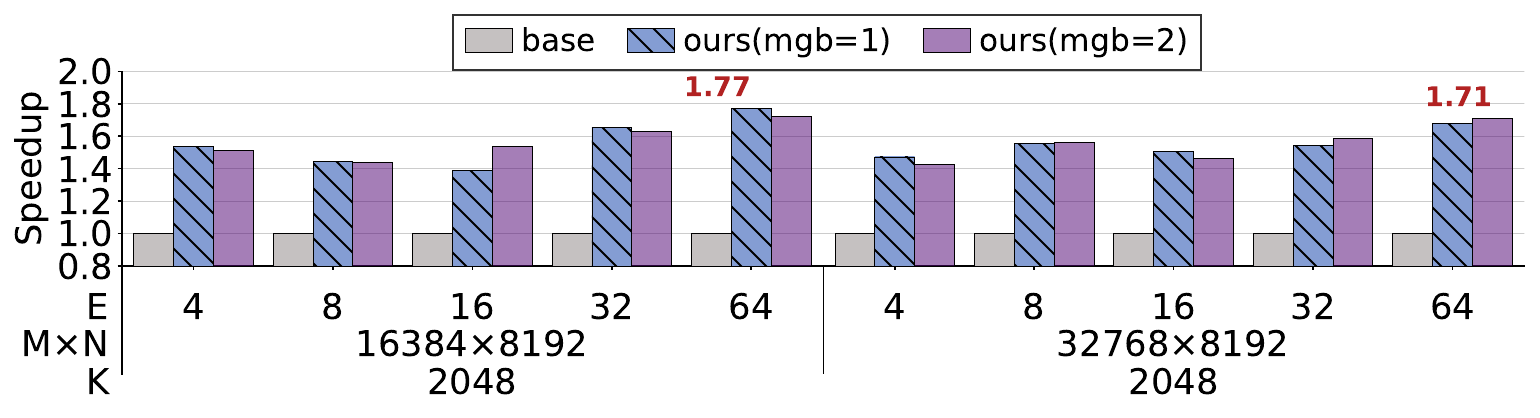}
\caption{stress\_skew}
\label{speedup_vs_fw_avg_total_max_ms_stress_skew_c8}}
\end{subfigure}
\caption{Speedup of the MoE layer.}
\label{e2e_base_ours}
\end{figure}
                   
\subsubsection{MoE Layer Performance}
Figure~\ref{e2e_base_ours} reports layer-level speedup over the baseline. Our approach delivers consistent gains across all router modes. The peak speedups are 1.78x for \texttt{balanced}, 1.73x for \texttt{moderate\_skew}, and 1.77x for \texttt{stress\_skew}.
The trend with $E$ stays similar to the operator-level observation.
Similarly, the MoE-layer speedups are smaller than the operator-level speedups because the non-operator stages contribute an unchanged cost and bound the layer-level gain. 

\subsubsection{Throughput} 
\begin{figure*}[t]
  \centering
  \includegraphics[width=\linewidth]{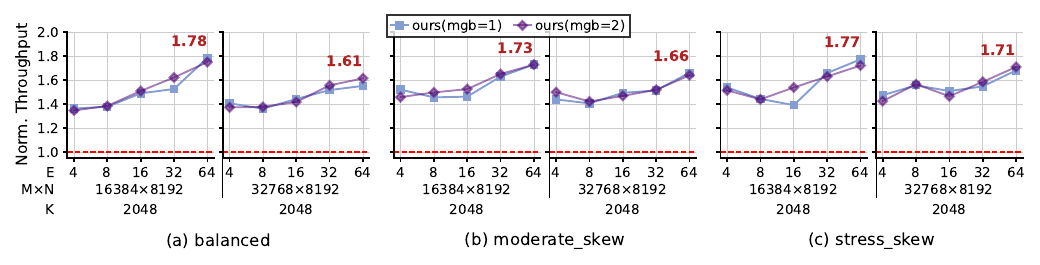}
  \caption{Normalized throughput (MTokens/sec).}
  \label{throughput_base_ours}
\end{figure*}

Figure~\ref{throughput_base_ours} shows the throughput of our approach (measured in millions of tokens per second, MTokens/sec), normalized to the baseline.
Our approach achieves peak throughput gains: 1.78x (\texttt{balanced}), 1.73x (\texttt{moderate\_skew}), and 1.77x (\texttt{stress\_skew}). 
We note similar trends for
the MoE-layer speedups above, since throughput is inversely related to latency. In practice, this means our latency savings translate directly into throughput gains, which is the metric that matters most in throughput-oriented MoE serving. 

\subsubsection{Resource Contention}\label{resource_contention}

\begin{figure*}[t]
\centering
\includegraphics[width=\linewidth]{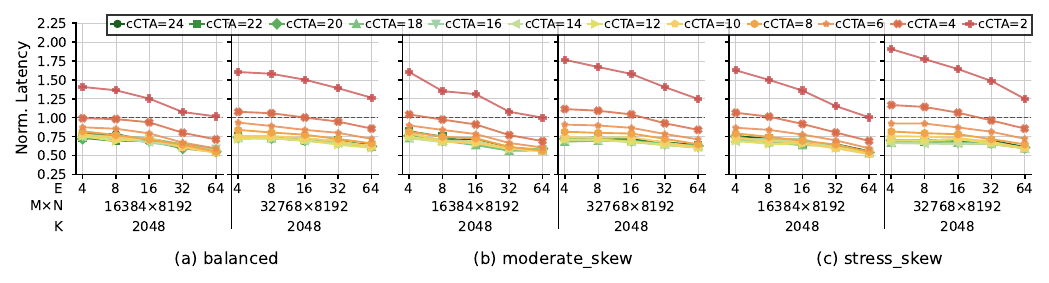}
\caption{Normalized latency ($\downarrow$ is better) of the MoE layer across all evaluated SM partitions.
The dashed line denotes the baseline.}
\label{contention_mgb1}
\end{figure*}

Resource contention is a common concern in GPU-based computation-communication overlap, where the computation and communication kernels share the same GPU resources, most notably the SM pool~\cite{Comet2025, jangda2022breaking, TileLink2025, eurosys26}.
A common intuition is to reserve as many SMs as possible for the compute-intensive GEMM and only a small number for communication kernels, since they are relatively compute-lightweight.
In practice, end-to-end performance is highly sensitive to the exact partition chosen. We analyze this sensitivity for our design in this section.
To expose the contention regime, we tune the number of consumer CTA ($cCTA$, one CTA per SM) from 2 to 24 with a stride of 2, 
as shown in Table~\ref{tbl:design_quality}. The remaining SMs run the GEMM kernel.

Figure~\ref{contention_mgb1} illustrates the MoE-layer latency of each configuration at $mgb = 1$, normalized to the baseline. 
Bars below 1.0 indicate that our overlap is faster than the baseline; bars above 1.0 indicate that it is slower. 
Over-constraining the consumer degrades performance sharply. The narrowest allocation at $cCTA = 2$ is slower than the baseline at every shape and router, reaching 1.91x the baseline under \texttt{stress\_skew} routing at $M = 32768$, $E = 4$. Two consumer SMs cannot drain completed tiles fast enough, and the resulting back-pressure collapses the overlap.
$cCTA = 4$ occasionally rises above $1.00$, especially under heavy load at $M = 32768$, $E = 4$ and $E = 8$.
The lowest latency is usually achieved when the consumer gets a moderate share of the SMs, with $cCTA$ in the range [10, 20]. 
We observe the similar trends for $mgb = 2$.

Router skew amplifies the contention penalty. The slowdown at $cCTA = 2$ is smallest under \texttt{balanced} routing, increases under \texttt{moderate\_skew}, and peaks under \texttt{stress\_skew}. 
As skew increases, a small subset of experts attracts most of the tokens, concentrating expert output production on the ranks that host them. 
The consumers on these heavily-loaded ranks face a sustained high-rate stream of ready tiles. With too few resources dedicated to communication, tiles cannot be issued fast enough, leaving a long communication tail after GEMM finishes. 
This demonstrates that an effective overlap method must be resource-aware, balancing compute throughput with communication progress rather than maximizing either side in isolation.
Production systems could benefit from an runtime-adaptive SM partition selector, which we leave to future work.

\subsection{Correctness Analysis}

We validate correctness of our approach by comparing the final per-token output of our implementation against the baseline on every 
configuration in Table~\ref{tbl:design_quality}, with identical seeds, weights, and routing decisions on both sides. 
The two paths share routing, first all-to-all, and scale
in Figure~\ref{overview_our_design}, and differ only in the overlapped expert compute and second all-to-all  where our path drives a persistent GEMM co-scheduled with a segment-granular all-to-all, while the baseline runs a per-expert GEMM followed by a bulk all-to-all. Any difference in the final per-token output comes solely from rounding noise inherent to using two different FP16 GEMM implementations on identical inputs. 
On every rank and every iteration, a check is considered as \textit{pass} when the relative distance between the two final per-token outputs falls within a tolerance of $8 \times 10^{-3}$. 
This tolerance sits several times above the FP16 rounding noise regime established for tensor-core dot-products~\cite{micikevicius2018mixed} and
far below what any structural correctness bug would produce.
Table~\ref{tab:correctness} summarizes 1440 independent correctness checks, covering 3 router distributions, 12 producer/consumer SM partitions, and 2 $mgb$ granularities across the shape grid in  Table~\ref{tbl:design_quality}. 
All checks pass except for one iteration on a single rank under \texttt{stress\_skew} that initially exceeded the tolerance. We repeated the identical configuration multiple times with the same inputs, and all reruns passed. Thus, the outlier was not reproduced in our following testing. So we conservatively treat it as a transient nondeterministic anomaly. 
The largest relative distance is $1.913 \times 10^{-3}$ under \texttt{stress\_skew}, roughly an order of magnitude below the tolerance and within expected FP16 rounding noise.

\begin{table}[t]
\centering
\small
\caption{Correctness of \textit{ours} vs \textit{base} output over configurations (\#Cfgs) in Table~\ref{tbl:design_quality}.
Tolerance: $\tau = 8 \times 10^{-3}$.}
\label{tab:correctness}
\begin{tabular}{lrrr}
\toprule
Router           & \#Cfgs & Max $|\varepsilon|_{\mathrm{rel}}$ & Pass \\
\midrule
balanced         & 480    & $1.875 \times 10^{-3}$ & 480/480 \\
moderate\_skew   & 480    & $1.884 \times 10^{-3}$ & 480/480 \\
stress\_skew     & 480    & $1.913 \times 10^{-3}$ & (480)/480 \\
\midrule
Total            & 1440    & $1.913 \times 10^{-3}$ & 1440/1440 \\
\bottomrule
\end{tabular}
\end{table}

\subsection{Overhead Analysis} 

The remote-owner-aligned row layout introduced in Section~\ref{sec:row_layout} inserts zero-filled rows at each owner-rank boundary so that the next group begins at a tile-aligned offset. These padded rows participate in GEMM computation but are excluded from the return transfer, so they impose a small compute overhead on the producer. 
The maximum number of padded rows per rank is $(W - 1)(tb_{M} - 1)$, where $W$ is the number of ranks.
For our 4-rank evaluation with tile heights up to $256$, this can potentially cause $765$ padding rows at most per rank, a few percent of the per-rank row count at problem sizes in Table~\ref{tbl:design_quality}. 
This overhead is theoretically smallest under \texttt{balanced} routing, where most owner-rank boundaries already land near $tb_{M}$ multiples, and largest under \texttt{stress\_skew}, where one owner-rank group may shrink and magnify its relative padding ratio. 
Despite this overhead, the benefits of our overlap outweigh this cost.

\section{Related work}\label{RW}

This section surveys prior works on overlap techniques to mitigate communication bottlenecks in multi-GPU systems.
Decomposition-based methods split the GEMM and its collective into smaller chunks, and then pipeline the chunks on separate streams.
CoCoNet~\cite{jangda2022breaking} and Wang et al.~\cite{ASPLOS2023_Wang} cast scheduling as a compiler problem and generate the kernel sequence automatically.
Chen et al.~\cite{ASPLOS24_Chen} build a hierarchical partition space with operator-, layer-, and model-level scheduling to maximize overlap efficiency in LLM training.
The technique has also been applied to LLM training in production~\cite{Domino2024, Aync_TP, MegaScale_USENIX}. 
Coarse-grained MoE pipelining methods~\cite{hwang2023tutel,he2022fastermoe, ScheMoE2024,PipeMoE_2023,MPipeMoE_2023} apply a similar chunking idea to MoE workloads by chunking the token dimension and pipelining expert compute with all-to-all communication. 
FasterMoE~\cite{he2022fastermoe} pipelines at a pipeline degree of two, and Tutel~\cite{hwang2023tutel} adds heuristic search over pipeline degrees. PipeMoE~\cite{PipeMoE_2023} and ScheMoE~\cite{ScheMoE2024} adapt the pipeline degree to the workload, with ScheMoE further scheduling MoE operators against intra- and inter-node bandwidth. MPipeMoE~\cite{MPipeMoE_2023} targets memory efficiency through adaptive pipeline parallelism.

Several studies have investigated fusion techniques that fuse computation and communication into a single kernel~\cite{Punniyamurthy_SC2024,Comet2025, FlashMoE,CCFuser2025}. 
These works aim to fuse or tightly integrate communication operations with computation phases to mitigate communication latency, improve data locality, and reduce synchronization costs.
Punniyamurthy et al.~\cite{Punniyamurthy_SC2024} study three operator-level fusion paradigms on AMD GPUs: embedding + all-to-all, GEMV + all-reduce, and GEMM + all-to-all.
For MoE specifically, Comet~\cite{Comet2025} restructures the shared tensor along the consumer-independent dimension, reschedules the GroupGEMM tile order, and fuses communication with computation in a thread-block-specialized kernel that adaptively partitions thread blocks between compute and communication.
CCFuser~\cite{CCFuser2025} replaces all-to-all operations with one-sided reads and writes using NVSHMEM, overlapping remote-data access with local-data compute through inter- and intra-thread-block scheduling inside fused kernels. 
FlashMoE~\cite{FlashMoE} packages the entire distributed MoE layer (token dispatch, expert GEMM, and combine) into a single CUDA kernel, taking the single-kernel philosophy further than the per-phase fusion of COMET and CCFuser. 
These fusion-based MoE systems achieve fine-grained overlap, but at the cost of intrusive software complexity and substantial per-target optimization.

\section{Conclusion\label{Conclusion}}

We presented a fine-grained computation-communication overlap design for MoE that hides the second all-to-all return transfer behind expert compute through tile-level signaling and scheduling.
Our approach relies on three co-design components: a remote-owner-aligned row layout that maps every tile to a single peer rank; a communication-aware tile schedule on a persistent rank-wide producer that emits remote-bound tiles first; and a consumer kernel that forwards completed tile segments to their owner ranks as soon as they are ready. 
Evaluated across three MoE models, multiple problem sizes, router distributions, and producer/consumer SM partitions, our design delivers substantial end-to-end speedup over both SoA MoE frameworks and conventional sequential implementations, with final per-token outputs that match the baseline up to FP16 rounding. 
Future work includes extension to the backward pass for training, adaptive SM partition selection, support for additional parallelism regimes, and deployment on larger machines.


\bibliographystyle{ACM-Reference-Format}
\bibliography{references}

@article{fedus2022switch,
  title={Switch Transformers: Scaling to Trillion Parameter Models with Simple and Efficient Sparsity},
  author={Fedus, William and Zoph, Barret and Shazeer, Noam},
  journal={Journal of Machine Learning Research},
  volume={23},
  number={120},
  pages={1--39},
  year={2022}
}

@inproceedings{micikevicius2018mixed,
title     = {Mixed Precision Training},
  author    = {Micikevicius, Paulius and Narang, Sharan and Alben, Jonah and Diamos, Gregory and Elsen, Erich and Garcia, David and Ginsburg, Boris and Houston, Michael and Kuchaiev, Oleksii and Venkatesh, Ganesh and Wu, Hao},
booktitle = {International Conference on Learning Representations (ICLR)},
year      = {2018},
archivePrefix = {arXiv},
url       = {https://arxiv.org/abs/1710.03740}
}

@online{llama4,
  author    = {Llama Team, AI @ Meta},
  title     = {The Llama 4 herd: The beginning of a new era of natively
multimodal AI innovation.},
  year      = {2025},
  url       = {https://ai.meta.com/blog/llama-4-multimodal-intelligence/},
  note      = {Accessed: 2025-12-12}
}

@inproceedings{eurosys26,
  author    = {Hong, Ke and Li, Xiuhong and Liu, Minxu and Mao, Qiuli and
               Wu, Tianqi and Huang, Zixiao and Chen, Lufang and
               Wang, Zhong and Zhang, Yichong and Zhu, Zhenhua and
               Dai, Guohao and Wang, Yu},
  title     = {Efficient and Adaptable Overlapping for Computation and
               Communication via Signaling and Reordering},
  booktitle = {European Conference on Computer Systems (EuroSys '26)},
  year      = {2026},
  note      = {arXiv:2504.19519},
}

@inproceedings{Comet2025,
title={{COMET}: Fine-grained Computation-communication Overlapping for Mixture-of-Experts},
author={Shulai Zhang and Ningxin Zheng and Haibin Lin and Ziheng Jiang and Wenlei Bao and Chengquan Jiang and Qi Hou and Weihao Cui and Size Zheng and Li-Wen Chang and Quan Chen and Xin Liu},
booktitle={Eighth Conference on Machine Learning and Systems (MLSys)},
year={2025},
}

@inproceedings{lepikhin2021gshard,
  title={{GShard}: Scaling Giant Models with Conditional Computation and Automatic Sharding},
  author={Lepikhin, Dmitry and Lee, HyoukJoong and Xu, Yuanzhong and Chen, Dehao and Firat, Orhan and Huang, Yanping and Krikun, Maxim and Shazeer, Noam and Chen, Zhifeng},
  booktitle={International Conference on Learning Representations (ICLR)},
  year={2021}
}

@inproceedings{dai2019transformerxl,
  title={Transformer-{XL}: Attentive Language Models Beyond a Fixed-Length Context},
  author={Dai, Zihang and Yang, Zhilin and Yang, Yiming and Carbonell, Jaime and Le, Quoc V. and Salakhutdinov, Ruslan},
  booktitle={Proceedings of the 57th Annual Meeting of the Association for Computational Linguistics (ACL)},
  pages={2978--2988},
  year={2019},
  publisher={Association for Computational Linguistics}
}

@inproceedings{jangda2022breaking,
  author    = {Jangda, Abhinav and Huang, Jun and Liu, Guodong and
               Sabet, Amir Hossein Nodehi and Maleki, Saeed and
               Miao, Youshan and Musuvathi, Madanlal and
               Mytkowicz, Todd and Saarikivi, Olli},
  title     = {Breaking the Computation and Communication Abstraction
               Barrier in Distributed Machine Learning Workloads},
  booktitle = {Proceedings of the International Conference on
               Architectural Support for Programming Languages and
               Operating Systems (ASPLOS)},
  year      = {2022},
}

@inproceedings{Lancet2024,
  author    = {Chenyu, Jiang and Ye, Tian and Zhen, Jia and Shuai, Zheng and Chuan, Wu and Yida, Wang},
  title     = {Lancet: Accelerating Mixture-of-Experts Training via Whole Graph Computation-Communication Overlapping},
  booktitle = {Proceedings of Machine Learning and Systems},
  pages = {74--86},
  volume = {6},
  year = {2024}
}

@inproceedings{ASPLOS24_Chen,
author = {Chen, Chang and Li, Xiuhong and Zhu, Qianchao and Duan, Jiangfei and Sun, Peng and Zhang, Xingcheng and Yang, Chao},
title = {Centauri: Enabling Efficient Scheduling for Communication-Computation Overlap in Large Model Training via Communication Partitioning},
year = {2024},
booktitle = {Proceedings of the 29th ACM International Conference on Architectural Support for Programming Languages and Operating Systems, Volume 3},
pages = {178–191},
numpages = {14},
location = {La Jolla, CA, USA},
series = {ASPLOS '24}
}

@INPROCEEDINGS{Punniyamurthy_SC2024,
  author={Punniyamurthy, Kishore and Hamidouche, Khaled and Beckmann, Bradford M.},
  booktitle={SC24: International Conference for High Performance Computing, Networking, Storage and Analysis}, 
  title={Optimizing Distributed ML Communication with Fused Computation-Collective Operations}, 
  year={2024},
  volume={},
  number={},
  pages={1-17},
  doi={10.1109/SC41406.2024.00094}}

@inproceedings{Aync_TP,
author = {PyTorch},
title = {Introducing Async Tensor Parallelism in PyTorch},
year = {2024},
url = {https://discuss.pytorch.org/t/distributed-w-torchtitan-introducing-async-tensor-parallelism-in-pytorch/209487},
}

@inproceedings{hwang2023tutel,
  title     = {Tutel: Adaptive Mixture-of-Experts at Scale},
  author    = {Hwang, Changho and Cui, Wei and Xiong, Yifan and Yang, Ziyue and
               Liu, Ze and Hu, Han and Wang, Zilong and Salas, Rafael and
               Jose, Jithin and Ram, Prabhat and Chau, HoYuen and Cheng, Peng and
               Yang, Fan and Yang, Mao and Xiong, Yongqiang},
  booktitle = {Proceedings of Machine Learning and Systems (MLSys)},
  volume    = {5},
  pages     = {269--287},
  year      = {2023}
}

@inproceedings{ScheMoE2024,
author = {Shi, Shaohuai and Pan, Xinglin and Wang, Qiang and Liu, Chengjian and Ren, Xiaozhe and Hu, Zhongzhe and Yang, Yu and Li, Bo and Chu, Xiaowen},
title = {ScheMoE: An Extensible Mixture-of-Experts Distributed Training System with Tasks Scheduling},
year = {2024},
isbn = {9798400704376},
url = {https://doi.org/10.1145/3627703.3650083},
booktitle = {Proceedings of the Nineteenth European Conference on Computer Systems},
pages = {236–249},
numpages = {14},
series = {EuroSys '24}
}

@INPROCEEDINGS{PipeMoE_2023,
  author={Shi, Shaohuai and Pan, Xinglin and Chu, Xiaowen and Li, Bo},
  booktitle={IEEE INFOCOM 2023 - IEEE Conference on Computer Communications}, 
  title={PipeMoE: Accelerating Mixture-of-Experts through Adaptive Pipelining}, 
  year={2023},
  volume={},
  number={},
  pages={1-10},
  doi={10.1109/INFOCOM53939.2023.10228874}}

@INPROCEEDINGS{MPipeMoE_2023,
  author={Zhang, Zheng and Yang, Donglin and Xia, Yaqi and Ding, Liang and Tao, Dacheng and Zhou, Xiaobo and Cheng, Dazhao},
  booktitle={2023 IEEE International Parallel and Distributed Processing Symposium (IPDPS)}, 
  title={MPipeMoE: Memory Efficient MoE for Pre-trained Models with Adaptive Pipeline Parallelism}, 
  year={2023},
  volume={},
  number={},
  pages={167-177}}

@inproceedings{CCFuser2025,
author = {Wang, Hulin and Xia, Yaqi and Yang, Donglin and Zhou, Xiaobo and Cheng, Dazhao},
title = {Harnessing Inter-GPU Shared Memory for Seamless MoE Communication-Computation Fusion},
year = {2025},
isbn = {9798400714436},
booktitle = {Proceedings of the 30th ACM SIGPLAN Annual Symposium on Principles and Practice of Parallel Programming},
pages = {170–182},
numpages = {13},
series = {PPoPP '25}
}

@inproceedings{FlashMoE,
  author    = {Osayamen Jonathan Aimuyo and Byungsoo Oh and Rachee Singh},
  title     = {{FlashMoE}: Fast Distributed {MoE} in a Single Kernel},
  booktitle = {Advances in Neural Information Processing Systems (NeurIPS '25)},
  year      = {2025},
  note      = {arXiv:2506.04667},
}

@inproceedings{EVT2024,
author = {Chen, Zhaodong and Kerr, Andrew and Cai, Richard and Kosaian, Jack and Wu, Haicheng and Ding, Yufei and Xie, Yuan},
title = {EVT: Accelerating Deep Learning Training with Epilogue Visitor Tree},
year = {2024},
isbn = {9798400703867},
publisher = {Association for Computing Machinery},
address = {New York, NY, USA},
url = {https://doi.org/10.1145/3620666.3651369},
doi = {10.1145/3620666.3651369},
booktitle = {Proceedings of the 29th ACM International Conference on Architectural Support for Programming Languages and Operating Systems, Volume 3},
pages = {301–316},
numpages = {16},
location = {La Jolla, CA, USA},
series = {ASPLOS '24}
}

@inproceedings{he2022fastermoe,
  author    = {He, Jiaao and Zhai, Jidong and Antunes, Tiago and
               Wang, Haojie and Luo, Fuwen and Shi, Shangfeng and
               Li, Qin},
  title     = {{FasterMoE}: Modeling and Optimizing Training of
               Large-Scale Dynamic Pre-Trained Models},
  booktitle = {Proceedings of the ACM SIGPLAN Symposium on Principles
               and Practice of Parallel Programming (PPoPP)},
  year      = {2022},
}

@online{cutlass,
  author = {{NVIDIA}},
  title  = {{CUTLASS}: {CUDA} Templates for Linear Algebra Subroutines},
  url    = {https://github.com/NVIDIA/cutlass},
  year   = {2023},
}

@inproceedings{T3_Pati_ASPLOS,
author = {Pati, Suchita and Aga, Shaizeen and Islam, Mahzabeen and Jayasena, Nuwan and Sinclair, Matthew D.},
title = {T3: Transparent Tracking \& Triggering for Fine-grained Overlap of Compute \& Collectives},
year = {2024},
isbn = {9798400703850},
url = {https://doi.org/10.1145/3620665.3640410},
doi = {10.1145/3620665.3640410},
booktitle = {Proceedings of the 29th ACM International Conference on Architectural Support for Programming Languages and Operating Systems (ASPLOS), Volume 2},
pages = {1146–1164},
numpages = {19},
}

@inproceedings{TileLink2025,
author={Size Zheng and Jin Fang and Xuegui Zheng and Qi Hou and Wenlei Bao and Ningxin Zheng and Ziheng Jiang and Dongyang Wang and Jianxi Ye and Haibin Lin and Li-Wen Chang and Xin Liu},
  title     = {TileLink: Generating Efficient Compute-Communication Overlapping Kernels using Tile-Centric Primitives},
  booktitle={Eighth Conference on Machine Learning and Systems (MLSys)},
  year      = {2025},
}

@article{Domino2024,
  author    = {Guanhua Wang and Chengming Zhang and Zheyu Shen and Ang Li and Olatunji Ruwase},
  title     = {Domino: Eliminating Communication in LLM Training via Generic Tensor Slicing and Overlapping},
  eprint    = {	arXiv:2409.15241 },
  year      = {2024},
}

@inproceedings{ASPLOS2023_Wang,
author = {Wang, Shibo and Wei, Jinliang and Sabne, Amit and Davis, Andy and Ilbeyi, Berkin and Hechtman, Blake and Chen, Dehao and Murthy, Karthik Srinivasa and Maggioni, Marcello and Zhang, Qiao and Kumar, Sameer and Guo, Tongfei and Xu, Yuanzhong and Zhou, Zongwei},
title = {Overlap Communication with Dependent Computation via Decomposition in Large Deep Learning Models},
year = {2022},
booktitle = {Proceedings of the 28th ACM International Conference on Architectural Support for Programming Languages and Operating Systems (ASPLOS)},
pages = {93–106},
numpages = {14}
}

@inproceedings{MegaScale_USENIX,
author = {Jiang, Ziheng and Lin, Haibin and Zhong, Yinmin and Huang, Qi and Chen, Yangrui and Zhang, Zhi and Peng, Yanghua and Li, Xiang and Xie, Cong and Nong, Shibiao and Jia, Yulu and He, Sun and Chen, Hongmin and Bai, Zhihao and Hou, Qi and Yan, Shipeng and Zhou, Ding and Sheng, Yiyao and Jiang, Zhuo and Xu, Haohan and Wei, Haoran and Zhang, Zhang and Nie, Pengfei and Zou, Leqi and Zhao, Sida and Xiang, Liang and Liu, Zherui and Li, Zhe and Jia, Xiaoying and Ye, Jianxi and Jin, Xin and Liu, Xin},
title = {MegaScale: scaling large language model training to more than 10,000 GPUs},
year = {2024},
isbn = {978-1-939133-39-7},
booktitle = {Proceedings of the 21st USENIX Symposium on Networked Systems Design and Implementation},
articleno = {41},
numpages = {16},
series = {NSDI'24}
}

@article{shoeybi2019megatron,
  author  = {Shoeybi, Mohammad and Patwary, Mostofa and Puri, Raul
             and LeGresley, Patrick and Casper, Jared and Catanzaro, Bryan},
  title   = {{Megatron-LM}: Training Multi-Billion Parameter Language Models
             Using Model Parallelism},
  journal = { arXiv:1909.08053},
  year    = {2019},
}

@misc{nvidia_grouped_gemm,
  author       = {{NVIDIA}},
  title        = {Grouped {GEMM} for {MoE}},
  howpublished = {\url{https://github.com/fanshiqing/grouped_gemm}},
  year         = {2024},
}

@misc{nvidia_te,
  author       = {{NVIDIA}},
  title        = {Transformer Engine},
  howpublished = {\url{https://github.com/NVIDIA/TransformerEngine}},
  year         = {2024},
}

\appendix









\end{document}